\documentclass{MYappolb}
\usepackage{epsfig}
\usepackage{cite}

\def\NPPS{Nucl. Phys. B (Proc. Suppl.)}
\def\PL{Phys. Lett.}

\def\PR{Phys. Rev.}

\def\MP{Int. J. Mod. Phys.}
\newcommand{\be}{\begin{equation}}
\newcommand{\ee}{\end{equation}}

\newcommand{\ba}{\begin{array}{c}}
\newcommand{\bat}{\begin{array}{cc}}
\newcommand{\ea}{\end{array}}
\newcommand{\beqn}{\begin{eqnarray}}
\newcommand{\eeqn}{\end{eqnarray}}

\newcommand{\bi}{\begin{itemize}}
\newcommand{\ei}{\end{itemize}}

\newcommand{\Br}{\mathrm{Br}}

\begin{document}
\title{SELECTED TOPICS ON TAU PHYSICS
\thanks{Invited talk at the XXXI International Conference of Theoretical Physics,
``Matter to the Deepest'', %%%: Recent Developments in Physics of Fundamental Interactions''
Ustro\'n, Poland, 5-11 September 2007}%
}
\author{A. Pich
\address{Departament de F\'{\i}sica Te\`orica, IFIC,
Universitat de Val\`encia--CSIC, \\  Apt. de Correus 22085, E-46071
Val\`encia, Spain}
%%%\and
%%%the Name(s) of other Author(s)
%%%\address{and their affiliation}
}
\maketitle
\begin{abstract}
The B Factories have generated a large amount of new results on the $\tau$ lepton.
The present status of some selected topics on $\tau$ physics is presented:
charged-current universality tests, bounds on lepton-flavour violation,
the determination of $\alpha_s$ from the inclusive $\tau$ hadronic width,
and the measurement of $|V_{us}|$ through the Cabibbo-suppressed decays
of the $\tau$ lepton.
\end{abstract}

\section{Introduction}

The known leptons provide clean probes to perform very precise tests
of the Standard Model and search for signals of new dynamics. The
electroweak gauge structure has been successfully tested at the
0.1\% to 1\% level, confirming the Standard Model framework
%%% as the modern theory of electroweak interactions
\cite{CERN07}.
Moreover, the hadronic $\tau$ decays turn out to be a beautiful
laboratory for studying strong interaction effects at low energies
\cite{taurev98,taurev06,taurev07,Stahl00}.
The $\tau$ is the only known lepton massive enough to decay into hadrons.
Its semileptonic decays are then ideally suited for studying the
hadronic weak currents in very clean conditions.
Accurate determinations of the QCD coupling, $|V_{us}|$ and the
strange quark mass have been obtained with $\tau$ decay data.

The huge statistics accumulated at the B Factories allow to explore
lepton-flavour-violating $\tau$ decay modes with increased
sensitivities beyond $10^{-7}$, which could be further pushed down
to few $10^{-9}$ at future facilities. Moreover, BESIII will soon
start taking data at the new Beijing Tau-Charm Factory. With the
excellent experimental conditions of the threshold region,
complementary information on the $\tau$ should be obtained, such as
an improved mass measurement.
Thus, $\tau$ physics is entering a new era, full of interesting
possibilities and with a high potential for new discoveries.

\section{Tests on charged-current universality}

In the Standard Model all lepton doublets have identical couplings
to the $W$ boson. Comparing the measured decay widths of leptonic or
semileptonic decays which only differ in the lepton flavour, one can
test experimentally that the $W$ interaction is indeed the same,
i.e. that \ $g_e = g_\mu = g_\tau \equiv g\, $. As shown in
Table~\ref{tab:ccuniv}, the present data verify the universality of
the leptonic charged-current couplings to the 0.2\% level.

%%%%%%%%%%%%%%%%%%%%  TABLES  %%%%%%%%%%%%%%%%%
\begin{table}[tbh]\centering
\caption{Experimental determinations of the ratios \ $g_l/g_{l'}$.}\vskip .1cm
\renewcommand{\arraystretch}{1.3}
\begin{tabular}{ccccc}
 \hline\hline &
%%% Present constraints on $|g_\tau/g_\mu|$
 $\Gamma_{\tau\to e}/\Gamma_{\mu\to e}$ &
 $\Gamma_{\tau\to\pi}/\Gamma_{\pi\to\mu}$ &
 $\Gamma_{\tau\to K}/\Gamma_{K\to\mu}$ &
 $\Gamma_{W\to\tau}/\Gamma_{W\to\mu}$
 \\ \hline
 %%%$\left|\frac{g_\tau}{g_\mu}\right|$
 $|g_\tau/g_\mu|$
 %%% & $1.0004\pm 0.0022$ & $0.996\pm 0.005$ & $0.979\pm 0.017$ & $1.039\pm 0.013$
 & $1.0006\; (22)$ & $0.996\; (5)$ & $0.979\; (17)$ & $1.039\; (13)$
 \\ \hline\hline &
%%% Present constraints on $|g_\mu/g_e|$
 $\Gamma_{\tau\to\mu}/\Gamma_{\tau\to e}$ &
 $\Gamma_{\pi\to\mu} /\Gamma_{\pi\to e}$ &
 $\Gamma_{K\to\mu} /\Gamma_{K\to e}$ &
 $\Gamma_{K\to\pi\mu} /\Gamma_{K\to\pi e}$
 \\ \hline
 $|g_\mu/g_e|$
 %%% & $1.0000\pm 0.0020$ & $1.0017\pm 0.0015$ & $1.012\pm 0.009$ & $1.0002\pm 0.0026$
 & $1.0000\; (20)$ & $1.0021\; (16)$ & $1.004\; (7)$ & $1.0021\; (25)$
 \\ \hline\hline &
%%% Present constraints on $|g_\mu/g_e|$ and $|g_\tau/g_e|$
 $\Gamma_{W\to\mu} /\Gamma_{W\to e}$ &
 \multicolumn{1}{||c}{} &
 $\Gamma_{\tau\to\mu}/\Gamma_{\mu\to e}$
 & $\Gamma_{W\to\tau}/\Gamma_{W\to e}$
 \\ \hline
 $|g_\mu/g_e|$
%%% & $0.997\pm 0.010$ &
%%% \multicolumn{1}{||c}{$|g_\tau/g_e|$} &
%%% $1.0004\pm 0.0023$ & $1.036\pm 0.014$\hfill
 & $0.997\; (10)$ & \multicolumn{1}{||c}{$|g_\tau/g_e|$}
 & $1.0005\; (23)$ & $1.036\; (14)$
 \\ \hline\hline
\end{tabular}
\label{tab:ccuniv}
\end{table}
%%%%%%%%%%%%%%%%%%%%%%%%%%%%%%%%%%%%%%%%%%%%%%%%\cite{Kl3}

The $\tau$ leptonic branching fractions and the $\tau$ lifetime are
known with a precision of $0.3\%$ \cite{taurev07}. A slightly improved lifetime
measurement could be expected from BaBar and Belle \cite{Lusiani}.
For comparison, the $\mu$ lifetime is already known with an accuracy of
$10^{-5}$, which should be further improved to $10^{-6}$ by the
MuLan experiment at PSI \cite{MuLan:07}.
The universality tests require also a good determination of
$m_\tau^5$, which is only known to the $0.06\%$ level \cite{PDG}. Two new
measurements of the $\tau$ mass have been published recently:
$$
m_\tau =\left\{ \begin{array}{lr} 1776.61\pm 0.13\pm
0.35~\mathrm{MeV}
&\; [\mathrm{Belle}],\\[10pt]
1776.81\, {}^{+\, 0.25}_{-\, 0.23} \pm 0.15~\mathrm{MeV} &\;
[\mathrm{KEDR}]. \ea\right.
$$
Belle \cite{BelleTauMass} has made a pseudomass analysis of
$\tau\to\nu_\tau 3\pi$ decays, while KEDR \cite{KEDR} measures the
$\tau^+\tau^-$ threshold production, taking advantage of a precise
energy calibration through the resonance depolarization method. In
both cases the achieved precision is getting close to the previous
BES-dominated value, $m_\tau = 1776.99\, {}^{+\, 0.29}_{-\, 0.26}$
\cite{PDG}. KEDR aims to obtain a final accuracy of 0.15 MeV. A
precision better than 0.1 MeV should be easily achieved at BESIII
\cite{MO}, through a detailed analysis of
$\sigma(e^+e^-\to\tau^+\tau^-)$ at threshold \cite{Pedro,Voloshin,SV:94}.

Table~\ref{tab:ccuniv} shows also the contraints obtained from
pion \cite{PDG} and kaon decays \cite{Kl3}, applying the
recently calculated radiative corrections at NLO in chiral perturbation
theory \cite{CR:07,CKNRT:02}. The accuracy achieved with $K_{l3}$ data is
already comparable to the one obtained from $\tau$ or $\pi_{l2}$ decays.

Owing to the limited statistics available, the decays $W^-\to l^-\nu_l$
only test universality at the 1\% level. At present,
$\Br(W\to\nu_\tau\tau)$ is $2.1\,\sigma/2.7\,\sigma$ larger than
$\Br(W\to \nu_e e / \nu_\mu\mu)$ \cite{LEPEWWG}. The stringent limits on
$|g_\tau/g_{e,\mu}|$ from $W$-mediated decays make unlikely that
this is a real effect.

\section{Lepton-flavour violating decays}

%%%%%%%%%%%%%%%%%%%%%%%%%%%%%%%%%%%%%%%%%%%%%%%%%%%%%%%%%%%%%%%%%%%
\begin{table}[tbh]\centering
\caption{Best limits (90\% C.L.) on
lepton-flavour-violating decays \cite{PDG,LFVbabar,LFVbelle}.}
\label{table:LFV}
\vspace{0.1cm}
\renewcommand{\arraystretch}{1.2} % enlarge line spacing
\begin{tabular}{ll|ll|ll}
%%%\begin{tabular}{cc|cc|cc}
\hline\hline
$\mu^-\to X^-$ & &&&&\\
 $e^-\gamma$ & $1.2\cdot 10^{-11}$ &
 $e^-2\gamma$ & $7.2\cdot 10^{-11}$ &
 $e^-e^-e^+$ & $1.0\cdot 10^{-12}$
 \\
 \hline\hline
 $\tau^-\to X^-$ & &&&&\\
 $e^-\gamma$ & $1.1\cdot 10^{-7}$ &
 $e^-e^+e^-$ & $3.6\cdot 10^{-8}$ &
 $e^-\mu^+\mu^-$ & $3.7\cdot 10^{-8}$
 \\
 $\mu^-\gamma$ & $4.5\cdot 10^{-8}$ &
 $\mu^-e^+e^-$ & $2.7\cdot 10^{-8}$ &
 $\mu^-\mu^+\mu^-$ & $3.2\cdot 10^{-8}$
 \\
 $e^-e^-\mu^+$ & $2.0\cdot 10^{-8}$ &
 $\mu^-\mu^-e^+$ & $2.3\cdot 10^{-8}$ &
 $\bar\Lambda\pi^-$ & $1.4\cdot 10^{-7}$
 \\
 $e^-\pi^0$ & $8.0\cdot 10^{-8}$ &
 $e^-\eta$ & $9.2\cdot 10^{-8}$ &
 $e^-\eta'$ & $1.6\cdot 10^{-7}$
 \\
 $\mu^-\pi^0$ & $1.1\cdot 10^{-7}$ &
 $\mu^-\eta$ & $6.5\cdot 10^{-8}$ &
 $\mu^-\eta'$ & $1.3\cdot 10^{-7}$
 \\
 $e^-\rho^0$ & $6.5\cdot 10^{-7}$ &
 $e^-\omega$ & $1.8\cdot 10^{-7}$ &
 $e^-\phi$ & $7.6\cdot 10^{-8}$
 \\
 $\mu^-\rho^0$ & $2.0\cdot 10^{-7}$ &
 $\mu^-\omega$ & $9.0\cdot 10^{-8}$ &
 $\mu^-\phi$ & $1.3\cdot 10^{-7}$
 \\
 $e^-K_S$ & $5.6\cdot 10^{-8}$ &
 $e^-K^{* 0}$ & $8.0\cdot 10^{-8}$ &
 $e^-\bar K^{* 0}$ & $7.7\cdot 10^{-8}$
 \\
 $\mu^-K_S$ & $4.9\cdot 10^{-8}$ &
 $\mu^-K^{*0}$ & $6.1\cdot 10^{-8}$ &
 $\mu^-\bar K^{*0}$ & $1.1\cdot 10^{-7}$
 \\
 $e^-K^+K^-$ & $1.4\cdot 10^{-7}$ &
 $e^-K^+\pi^-$ & $1.6\cdot 10^{-7}$ &
 $e^-\pi^+K^-$ & $3.2\cdot 10^{-7}$
 \\
 $\mu^-K^+K^-$ & $2.5\cdot 10^{-7}$ &
 $\mu^-K^+\pi^-$ & $3.2\cdot 10^{-7}$ &
 $\mu^-\pi^+K^-$ & $2.6\cdot 10^{-7}$
 \\
 $e^-\pi^+\pi^-$ & $1.2\cdot 10^{-7}$ &
 $\mu^-\pi^+\pi^-$ & $2.9\cdot 10^{-7}$ &
 $\Lambda\pi^-$ & $7.2\cdot 10^{-8}$
 \\
 $e^+\pi^-\pi^-$ & $2.0\cdot 10^{-7}$ &
 $e^+K^-K^-$ & $1.5\cdot 10^{-7}$ &
 $e^+\pi^-K^-$ & $1.8\cdot 10^{-7}$
 \\
 $\mu^+\pi^-\pi^-$ & $0.7\cdot 10^{-7}$ &
 $\mu^+K^-K^-$ & $4.4\cdot 10^{-7}$ &
 $\mu^+\pi^-K^-$ & $2.2\cdot 10^{-7}$
 \\ \hline\hline
\end{tabular}\end{table}
%%%%%%%%%%%%%%%%%%%%%%%%%%%%%%%%%%%%%%%%%%%%%%%%%%%%%%%%%%%%%%%%%%%

We have now clear experimental evidence that neutrinos are
massive particles and there is mixing in the lepton sector.
The smallness of neutrino masses implies a strong suppression of
neutrinoless lepton-flavour-violating processes, which can be
avoided in models with other sources of lepton flavour violation,
not related to $m_{\nu_i}$.
The scale of the flavour-violating new-physics interactions can be
constrained imposing the requirement of a viable leptogenesis.
Recent studies within different new-physics scenarios find interesting
correlations between $\mu$ and $\tau$ lepton-flavour-violating decays,
with $\mu\to e\gamma$ often expected to be close to the present exclusion limit
\cite{CIP:07,GCIW:07,AAHT:06,AH:06,BBDPT:07,CMPSVV:07,PA:06}.

The B Factories are pushing the experimental
limits on neutrinoless lepton-flavour-violating
$\tau$ decays beyond the $10^{-7}$ level \cite{LFVbabar,LFVbelle},
increasing in a drastic way the sensitivity to new physics scales.
Future experiments could push further some limits to the $10^{-9}$
level \cite{SuperB}, allowing to explore interesting and
totally unknown phenomena. Complementary information will be
provided by the MEG experiment, which will search for $\mu^+\to
e^+\gamma$ events with a sensitivity of $10^{-13}$ \cite{MEG}.
There are also ongoing projects at J-PARC aiming to study $\mu\to e$
conversions in muonic atoms, at the $10^{-16}$ \cite{PRISM-I}
or even $10^{-18}$ \cite{PRISM-II} level.

\section{The inclusive hadronic width of the tau lepton}

The hadronic decays of the $\tau$ lepton provide a very clean laboratory
to perform precise tests of the Standard Model \cite{taurev07}.
The inclusive character of the total $\tau$ hadronic width renders
possible an accurate calculation of the ratio
\cite{BR:88,NP:88,BNP:92,LDP:92a,QCD:94}
%[$(\gamma)$ represents additional photons or lepton pairs]
%
\begin{equation}\label{eq:r_tau_def}
 R_\tau \,\equiv\, { \Gamma [\tau^- \to \nu_\tau
 \,\mathrm{hadrons}\, (\gamma)] \over \Gamma [\tau^- \to \nu_\tau e^-
 {\bar \nu}_e (\gamma)] }\, = \, R_{\tau,V} + R_{\tau,A} + R_{\tau,S}\, ,
\end{equation}
using analyticity constraints and the Operator Product Expansion.
One can separately compute the contributions associated with
specific quark currents:
$R_{\tau,V}$ and $R_{\tau,A}$ correspond to the Cabibbo-allowed
decays through the vector and axial-vector currents, while
$R_{\tau,S}$ contains the remaining Cabibbo-suppressed
contributions.

The theoretical prediction for $R_{\tau,V+A}$ can be expressed as
\cite{BNP:92}
\begin{equation}\label{eq:Rv+a}
 R_{\tau,V+A} \, =\, N_C\, |V_{ud}|^2\, S_{\mathrm{EW}} \left\{ 1 +
 \delta_{\mathrm{P}} + \delta_{\mathrm{NP}} \right\} ,
\end{equation}
where $N_C=3$ is the number of quark colours
and $S_{\mathrm{EW}}=1.0201\pm 0.0003$ contains the
electroweak radiative corrections \cite{MS:88,BL:90,ER:02}.
The dominant correction ($\sim 20\%$) is the perturbative QCD
contribution $\delta_{\mathrm{P}}$, which is fully known to
$O(\alpha_s^3)$ \cite{BNP:92} and includes a resummation of the most
important higher-order effects \cite{LDP:92a,PI:92}.

Non-perturbative contributions are suppressed by six powers of the
$\tau$ mass \cite{BNP:92} and, therefore, are very small. Their
numerical size has been determined from the invariant-mass
distribution of the final hadrons in $\tau$ decay, through the study
of weighted integrals \cite{LDP:92b},
\begin{equation}
 R_{\tau}^{kl} \,\equiv\, \int_0^{m_\tau^2} ds\, \left(1 - {s\over
 m_\tau^2}\right)^k\, \left({s\over m_\tau^2}\right)^l\, {d
 R_{\tau}\over ds} \, ,
\end{equation}
which can be calculated theoretically in the same way as $R_{\tau}$.
The predicted suppression \cite{BNP:92} of the non-perturbative
corrections has been confirmed by ALEPH \cite{ALEPH:05}, CLEO
\cite{CLEO:95} and OPAL \cite{OPAL:98}. The most recent analysis
\cite{ALEPH:05} gives
\begin{equation}\label{eq:del_np}
 \delta_{\mathrm{NP}} \, =\, -0.0043\pm 0.0019 \, .
\end{equation}
%

%%%%%%%%%%%%%%%%%% FIGURE %%%%%%%%%%%%%%%%%%%%%
%              alpha_s running
%
\begin{figure}[tbh]
\label{fig:alpha_s} \centering
\includegraphics[width=7.6cm]{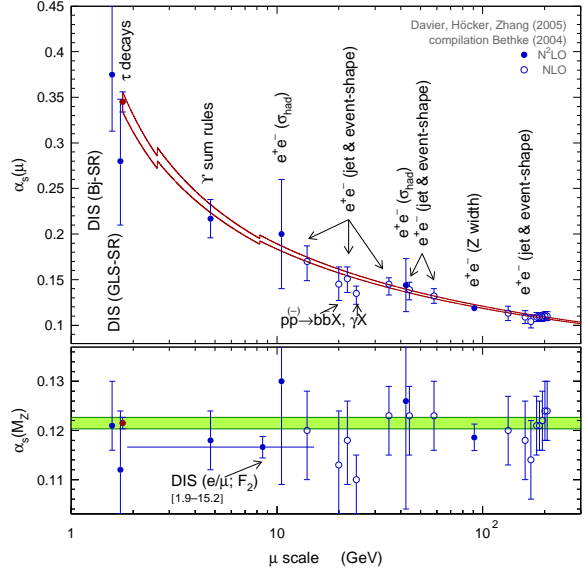}
%%%\vspace{-0.5cm}
\caption{Measured values of $\alpha_s$ at different
scales. The curves show the energy dependence predicted by QCD,
using $\alpha_s(m_\tau^2)$ as input. The corresponding extrapolated
$\alpha_s(M_Z^2)$ values are shown at the bottom, where the shaded
band displays the $\tau$ decay result within errors \cite{DHZ:05}.}
\end{figure}
%%%%%%%%%%%%%% END FIGURE %%%%%%%%%%%%%%%%%%%%%

The QCD prediction for $R_{\tau,V+A}$ is then completely dominated
by the perturbative contribution; non-perturbative effects being
smaller than the perturbative uncertainties from uncalculated
higher-order corrections. The result turns out to be very sensitive
to the value of $\alpha_s(m_\tau)$, allowing for an accurate
determination of the fundamental QCD coupling \cite{NP:88,BNP:92}.
The experimental measurement $R_{\tau,V+A}= 3.471\pm0.011$ implies
\cite{DHZ:05}
\begin{equation}\label{eq:alpha}
 \alpha_s(m_\tau)  \,=\,  0.345\pm
 0.004_{\mathrm{exp}}\pm 0.009_{\mathrm{th}} \, .
\end{equation}

The strong coupling measured at the $\tau$ mass scale is
significantly larger than the values obtained at higher energies.
From the hadronic decays of the $Z$, one gets $\alpha_s(M_Z) =
0.1186\pm 0.0027$ \cite{LEPEWWG}, which differs from $\alpha_s(m_\tau)$
by more than twenty standard deviations. After
evolution up to the scale $M_Z$ \cite{Rodrigo:1998zd}, the strong
coupling constant in (\ref{eq:alpha}) decreases to \cite{DHZ:05}
\begin{equation}\label{eq:alpha_z}
 \alpha_s(M_Z)  \, =\,  0.1215\pm 0.0012 \, ,
\end{equation}
in excellent agreement with the direct measurements at the $Z$ peak
and with a similar accuracy. The comparison of these two
determinations of $\alpha_s$ in two very different energy regimes, $m_\tau$
and $M_Z$, provides a beautiful test of the predicted running of the
QCD coupling; i.e., a very significant experimental verification of
{\it asymptotic freedom}.

\section{$|V_{us}|$ determination from Cabibbo-suppressed tau decays}

The separate measurement of the $|\Delta S|=0$ and $|\Delta S|=1$
tau decay widths provides a very clean determination of $V_{us}\,$
\cite{GJPPS:05,PI:07b}.
To a first approximation the Cabibbo mixing can be directly obtained
from experimental measurements, without any theoretical input.
Neglecting the small SU(3)-breaking corrections from the $m_s-m_d$
quark-mass difference, one gets:
\begin{equation}
 |V_{us}|^{\mathrm{SU(3)}}\, =\; |V_{ud}| \;\left(\frac{R_{\tau,S}}{R_{\tau,V+A}}\right)^{1/2}
 =\; 0.210\pm 0.003\qquad [0.215\pm 0.003]\, .
\end{equation}
We have used $|V_{ud}| = 0.97377\pm 0.00027$ \cite{PDG},
$R_\tau = 3.640\pm 0.010$       %%%\equiv R_{\tau,V+A}+R_{\tau,S} = 3.640\pm 0.010$
and the value $R_{\tau,S}=0.1617\pm 0.0040$ \cite{PI:07b}, which results from the
most recent BaBar \cite{BA:07} and Belle \cite{BE:07} measurements of Cabibbo-suppressed 
tau decays \cite{Banerjee}.
%%%\cite{BA:07,BE:07}.
The new branching ratios measured by BaBar and Belle are all smaller than the previous
world averages, which translates into a smaller value of $R_S$ and $|V_{us}|$.
For comparison, we give in brackets the result obtained with the previous
value $R_{\tau,S}=0.1686\pm 0.0047$ \cite{DHZ:05}.

This rather remarkable determination is only slightly shifted by
the small SU(3)-breaking contributions induced by the strange quark mass.
These corrections can be theoretically estimated through a QCD analysis of the differences
\cite{GJPPS:05,PI:07b,Davier,PP:99,ChDGHPP:01,ChKP:98,KKP:01,MW:06,KM:00,MA:98,BChK:05}
\begin{equation}
 \delta R_\tau^{kl}  \,\equiv\,
 {R_{\tau,V+A}^{kl}\over |V_{ud}|^2} - {R_{\tau,S}^{kl}\over |V_{us}|^2}\, .
\end{equation}
Since the strong interactions are flavour blind, these quantities vanish in the SU(3) limit.
The only non-zero contributions are proportional to powers of
the quark mass-squared difference $m_s^2(m_\tau)-m_d^2(m_\tau)$ or to vacuum expectation
values of SU(3)-breaking operators such as $\delta O_4
\equiv \langle 0|m_s\bar s s - m_d\bar d d|0\rangle \approx (-1.4\pm 0.4)
\cdot 10^{-3}\; \mathrm{GeV}^4$ \cite{PP:99,GJPPS:05}. The dimensions of these operators
are compensated by corresponding powers of $m_\tau^2$, which implies a strong
suppression of $\delta R_\tau^{kl}$ \cite{PP:99}:
\begin{equation}\label{eq:dRtau}
 \delta R_\tau^{kl}\,\approx\,  24\, S_{\mathrm{EW}}\,\left\{ {m_s^2(m_\tau)\over m_\tau^2} \,
 \left( 1-\epsilon_d^2\right)\,\Delta_{kl}(\alpha_s)
 - 2\pi^2\, {\delta O_4\over m_\tau^4} \, Q_{kl}(\alpha_s)\right\}\, ,
\end{equation}
where $\epsilon_d\equiv m_d/m_s = 0.053\pm 0.002$ \cite{LE:96}.
The perturbative QCD corrections $\Delta_{kl}(\alpha_s)$ and
$Q_{kl}(\alpha_s)$ are known to $O(\alpha_s^3)$ and $O(\alpha_s^2)$,
respectively \cite{PP:99,BChK:05}.

The theoretical analysis of $\delta R_\tau\equiv\delta R_\tau^{00}$
 involves the two-point correlation
functions of vector and axial-vector quark currents, which can
be separated into their transverse ($J=1$) and longitudinal ($J=0$) components.
The longitudinal contribution to $\Delta_{00}(\alpha_s)$ shows a rather
pathological behaviour, with clear signs of being a non-convergent perturbative
series. Fortunately, the corresponding longitudinal contribution to
$\delta R_\tau$ can be estimated phenomenologically with a much better
accuracy, $\delta R_\tau|^{L}\, =\, 0.1544\pm 0.0037$ \cite{GJPPS:05,JOP:06},
because it is dominated by far by the well-known $\tau\to\nu_\tau\pi$
and $\tau\to\nu_\tau K$ contributions. To estimate the remaining transverse
component, one needs an input value for the strange quark mass. Taking the
range
$m_s(m_\tau) = (100\pm 10)\:\mathrm{MeV}$ \
[$m_s(2\:\mathrm{GeV}) = (96\pm 10)\:\mathrm{MeV}$],
which includes the most recent determinations of $m_s$ from QCD sum rules
and lattice QCD \cite{JOP:06},
one gets finally $\delta R_{\tau,th} = 0.216\pm 0.016$, which implies \cite{PI:07b}
\be\label{eq:Vus_det}
 |V_{us}| \,=\, \left(\frac{R_{\tau,S}}{\frac{R_{\tau,V+A}}{|V_{ud}|^2}-\delta
 R_{\tau,\mathrm{th}}}\right)^{1/2}
 \; =\; \left\{
 \ba
 \, 0.2165\pm 0.0026_{\mathrm{\, exp}}\pm 0.0005_{\mathrm{\, th}}
 \\[15pt]
 \left[0.2212\pm 0.0031_{\mathrm{\, exp}}\pm 0.0005_{\mathrm{\, th}}\right]
 \ea\right.\, .
\ee
Again, the first number is the updated value, including the recent BaBar and Belle data,
while the number inside brackets gives the previous result.

Sizeable changes on the experimental determination of $R_{\tau,S}$ are to be expected from
the full analysis of  the huge BaBar and Belle data samples. In particular, the high-multiplicity
decay modes are not well known at present and their effect has been just roughly estimated or
simply ignored. Thus, the result (\ref{eq:Vus_det}) could easily fluctuate in the near future.
However, it is important to realize that the final error of the $V_{us}$ determination from
$\tau$ decay is completely dominated by the experimental uncertainties. If $R_{\tau,S}$
is measured with a 1\% precision, the resulting $V_{us}$ uncertainty will
get reduced to around 0.6\%, i.e. $\pm 0.0013$, making $\tau$ decay the best source of
information about $V_{us}$.

An accurate experimental measurement of the invariant-mass distribution of the final
hadrons in Cabibbo-suppressed $\tau$ decays could make possible a simultaneous determination
of $V_{us}$ and the strange quark mass, through a correlated analysis of
several weighted differences $\delta R_\tau^{kl}$. The extraction of $m_s$ suffers from
theoretical uncertainties related to the convergence of the perturbative series
$\Delta_{kl}^{L+T}(\alpha_s)$, which makes necessary a better
understanding of these corrections. Further work in this direction is in progress
\cite{GJPPS:07}.

%%%%%%%%%%%%%%%%%%%%%%%%%%%%%%%%%%%%%%%%%%%%%%%%%%%%%%%%%%%%%%%%%%%%%%%%

\section*{Acknowledgements}
I want to thank Michal Czakon, Henryk Czy\.{z} and Janusz Gluza for organizing
an enjoyable conference. This work has been supported by
the Spanish Ministry of Education (Grant FPA2004-00996),
by Generalitat Valenciana (GVACOMP2007-156), by EU funds for regional development
and by the EU MRTN-CT-2006-035482 (FLAVIA{\it net}).

%%%%%%%%%%%%%%%%%%%%%%%%%%%%%%%%%%%%%%

\end{document}